# Entanglement between states of single quadrupole nuclear spin

G. B. Furman[1,2], V. M. Meerovich[1], and V. L. Sokolovsky[1]

[1]*Physics Department, Ben Gurion University,*

*POB 653, Beer Sheva, 84105, Israel and*

[2]*Ohalo College, POB 222, Katsrin, 12900, Israel*

## Abstract

We study entanglement between quantum states of multi level spin system of a single particle considering a nucleus with spin 3/2 in both the internal electric field gradient and the external magnetic field. It was shown that entanglement is achieved by applying a magnetic field to a single particle at low temperature ( 5 mK). In this temperature range, the numerical calculation revealed the coincidence between magnetization and concurrence. As a result, the magnetization can be used as an entanglement witness for such systems.

## I. INTRODUCTION

Quantum systems in an entangled state [1–3] can be used as a primary quantum information channel to perform computational [5], communicational [6], metrological [7, 8] and cryptographic tasks that are impossible for classical systems. These possible applications of entangled quantum states stimulate intensive research in the fields of generation and manipulation of them [9]. Entangled states may appear in various systems of interacting quantum particles, such as phonons, ions, electron, nuclear spins, and of a single particle interacting with environment [11–17] and in systems of interacting qubits [1–3]. It should be emphasized that property of entanglement can be considered regardless of the nature of qubits[10, 11].

One of the most intensively investigated systems is clusters of coupling nuclear spins [1–3] which received considerable attention as a platform for the practical implementation of a quantum computer (QC) by using nuclear magnetic resonance (NMR) technique [18–21]. The basic idea for an NMR QC is that two stationary states of spin $\frac{1}{2}$ in an applied magnetic field represent naturally one qubit. To perform complex quantum computations, operation with a set of quantum gates has to be realized using several coupling spins $\frac{1}{2}$. The strength of the coupling (such as dipole-dipole, scalar or exchange interactions) between different spins and interaction of the spins with environment determine the decoherence time. Short decoherence times limit possible calculation times of the QC [22]. In order to overcome this weak point, it has been proposed to construct qubits using spins greater than $\frac{1}{2}$ in a strong magnetic field, thus eliminating the requirement of interaction between spins and decreasing interaction of qubits with environment [23–29]

Factorization of the spin system of a quadrupole nucleus into subsystems has been proposed at the dawn of nuclear quadrupole resonance (NQR) [30, 31]. It was shown that a spin $\frac{3}{2}$ system is equivalent to a system of two magnetization vectors [30, 31] and can be represented using the Pauli spin matrices 2x2 [32]. It means that a single spin 3/2 is isomorphic to a system consists of two dipolar coupling spins 1/2, which can be considered as qubits. This was experimentally confirmed by using NQR technique with quadrupole splitting [25, 26]. The feasibility of quantum computing based on a pure (without external magnetic fields) NQR technique was theoretically investigated in detail in [27]. Using the resonance excitation technique and the level-crossing method, it was proposed three quan-tum logic gates: a controlled $NOT$, $SWAP$ and $NOT_2$ . Thus the method to synthesize qubits from a set of the spin states of a single particle with spin higher 1/2 has been developed. It is logical to raise the question of entanglement of these qubits. The need of solving this problem is connected with the increased interest, on the one hand, in application of single-particle entanglement, for example, in quantum cryptography [12], and, on the other hand, in protection from decoherence of the entangled states. There are only few papers which deal with entanglement between quantum states of a single quantum particle [12–17]. Single-particle entanglement was investigated for the following quantum systems: a superposition of the vacuum state and a single-photon state [12, 14], one atom distributed over two spatially distinct traps [17], a single photon with an orbital angular momentum $l = 1$, an electric dipole photon with a total angular momentum $j = 1$, and the superfluid $^3He$ [15].

Our present purpose is to investigate entanglement between the quantum states of a multi-level spin system of a single particle. We consider a nucleus with spin $\frac{3}{2}$ being in an internal electric field gradient (EFG) and an external magnetic field when the quadrupole interaction energy is of the order of the magnitude or even greater than the Zeeman one. Results of computer simulations of entanglement vs. temperature and magnetic field are presented for real spin systems with a non-equadistant energy spectrum.

## II. NQR IN MAGNETIC FIELD

Let us consider a system of spins $I$ ($I > 1/2$) in the thermodynamic equilibrium with the density matrix

$$\rho = \mathbb{Z}^{-1} \exp\left(-\frac{\mathcal{H}}{k_B T}\right). \qquad (1)$$

Here $k_B$ is the Boltzmann constant, $T$ is the lattice temperature, $\mathbb{Z} = Tr\left\{\exp\left(-\frac{\mathcal{H}}{k_B T}\right)\right\}$ is the partition function. In the general case, the Hamiltonian $\mathcal{H}$ can consist of the Zeeman ($\mathcal{H}_M$) and the quadrupole ($\mathcal{H}_Q$) parts,

$$\mathcal{H} = \mathcal{H}_M + \mathcal{H}_Q. \qquad (2)$$

The Zeeman interaction between the applied magnetic field, $\vec{H}_0$ and nuclear spin is the external factor for a crystal and the direction of this field is chosen as the $z$-axis of the laboratory frame, $\vec{H}_0 = H_0 \vec{z}$, where $H_0$ is the strength of the external magnetic field. The part of the Hamiltonian describing this interaction is

$$\mathcal{H}_M = -\gamma H_0 I_z, \tag{3}$$

where $\gamma$ is the gyromagnetic ratio of the nucleus with spin $I$, $I_z$ is the projection of the individual spin angular momentum operators $\vec{I}$ on the $z$- axis.

Quadrupole coupling exists between a non-spherical nuclear charge distribution and an electric field gradient (EFG) generated by nuclear surroundings. It is possible to reduce the EFG symmetric tensor to a diagonal form by finding the principal axes frame (PAF) with the $Z$- and $X$-axises directed along the maximum and minimum of EFG, respectively, $|V_{ZZ}| \geq |V_{YY}| \geq |V_{XX}|$, where $V_{\xi\xi} = \frac{\partial^2 V}{\partial \xi^2}$ ($\xi = X, Y, Z$) and $V$ is the potential of the electric field. In the laboratory frame the quadrupolar Hamiltonian part as a scalar product of an irreducible spherical tensor of the second-rank $V_m$ and second-rank tensor operator $Q_m$ can be presented in the following form [33–35]

$$H_Q = \frac{eQq_{ZZ}}{4I(2I-1)} \sum_{m=-2}^{m=2} (-1)^m V_{-m} Q_m \tag{4}$$

where $eQq_{ZZ}$ is the quadrupole coupling constant of EFG, the quadrupole tensor elements are taken as follows:

$$Q_0 = \frac{1}{2} \left( 3I_z^2 - \vec{I}^2 \right),$$
$$Q_{\pm 1} = \pm \frac{1}{2} \left( I_z I_{\pm} + I_{\pm} I_z \right),$$
$$Q_{\pm 2} = \frac{1}{2} I_{\pm}^2 . \tag{5}$$

$I_{\pm}$ are the raising and lowering operators of the spin, and

$$V_0 = \left[ \left( 3\cos^2 \theta - 1 \right) + \eta \cos 2\varphi \sin^2 \theta \right],$$
$$V_{\pm 1} = \pm \left[ \eta \sin \theta \left( \cos \theta \cos 2\varphi \pm i \sin 2\varphi \right) + \frac{3}{2} \sin 2\theta \right],$$
$$V_{\pm 2} = \left[ \frac{3}{2} \sin^2 \theta + \eta \cos 2\varphi \left( 1 + \cos^2 \theta \right) \mp \frac{i}{2} \eta \sin 2\varphi \cos \theta \right]. \tag{6}$$

Here $\theta$ and $\varphi$ refer to the polar and azimuthal angles determining the orientation of the laboratory frame $z$-axis in the PAF coordinate system. The asymmetry parameter $\eta$ is defined as $\eta = \frac{V_{YY} - V_{XX}}{V_{ZZ}}$, which may vary between 0 and 1.

Using the Hamiltonian (2) the density matrix (1) can be represented as a function of parameters $\alpha = \frac{\gamma H_0}{k_B T}$ and $\beta = \frac{eQq_{ZZ}}{4I(2I-1)k_B T}$:

$$\rho = \mathbb{Z}^{-1} \exp\left(-\alpha I_z - \beta \sum_{m=-2}^{m=2} (-1)^m V_{-m} Q_m\right). \tag{7}$$

The density matrix (7) can be employed to obtain information on the dependence of entanglement on the magnetic field, quadrupole coupling constant, orientation of the crystal principal axis in the laboratory frame, and temperature.

Below we consider entanglement in a system of spins 3/2. A suitable system for studying by NQR technique is a high temperature superconductor $YBa_2Cu_3O_{7-\delta}$ containing the $^{63}Cu$ and $^{65}Cu$ nuclei with spin $\frac{3}{2}$ possessing quadrupole moments $Q = -0.211 \times 10^{-24}$ cm$^2$ and $-0.195 \times 10^{-24}$ cm$^2$, respectively [36]. There are two different locations of copper ions in this structure: the first are the copper ion sites at the center of an oxygen rhombus-like plane while the second one is five-coordinated by an apically elongated rhombic pyramid. The four-coordinated copper ion site, EFG is highly asymmetric ($\eta \geq 0.92$) while the five-coordinated copper ion site, EFG is nearly axially symmetric ($\eta = 0.14$). The quadrupole coupling constant ($eQq_{ZZ}$) of $^{63}Cu$ in the four-coordinated copper ion site is 38.2 MHz  and in the five-coordinated copper ion site is 62.8 MHz [36].

## III.  CONCURRENCE BETWEEN STATES OF THE NUCLEAR SPIN

In order to quantify entanglement, the concurrence $C$ is usually used [37]. For the maximally entangled states, the concurrence is $C = 1$, while for the separable states $C = 0$. The concurrence of a quantum system with the density matrix presented in the Hilbert space as a matrix $4 \times 4$ is expressed by the formula [37]:

$$C = \max\left\{0, 2\nu - \sum_{i=1}^{4} \nu_i\right\} \tag{8}$$

where $\nu = \max\{\nu_1, \nu_2, \nu_3, \nu_4\}$ and $\nu_i$ $(i = 1, 2, 3, 4)$ are the square roots of the eigenvalues of the product

$$R = \rho\tilde{\rho} \tag{9}$$

with

$$\tilde{\rho} = G\bar{\rho}G \tag{10}$$

where $\bar{\rho}(\alpha, \beta)$ is the complex conjugation of the density matrix (7) taken in the $I_z$ representation and

$$G = \begin{pmatrix} 0 & 0 & 0 & -1 \\ 0 & 0 & 1 & 0 \\ 0 & 1 & 0 & 0 \\ -1 & 0 & 0 & 0 \end{pmatrix}. \tag{11}$$

The numerical simulations of entanglement of the spin states are performed using the software based on the *Mathematica* package. In the equilibrium the states of the system determined by Eq. (7) are separable without applying external magnetic filed ($\alpha = 0$) at any temperature. At large temperature ($\beta < 1$) and low magnetic field strength ($\alpha < 1$) the concurrence is zero. Entanglement appears in the course of temperature decrease and increasing the magnetic field.

The concurrence depends on the orientation of the PAF frame relatively the laboratory frame (Fig. 1). As an example we present the concurrence as a function of the angels at $\alpha = \beta = 5$. When the z- and Z-axes are parallel, $\theta = 0$ or $\pi$ entanglement is absent. At the asymmetry parameters $\eta = 0$ the concurrence is independent of the angle $\varphi$ and there are two equaled maximums of $C = 0.21$ at $\theta = 0.94$ and 2.20 (Fig.1a). With an increase of the asymmetry parameters the concurrence dependence on the angle $\varphi$ appears (Fig. 1b and 1c). At the low asymmetry parameters, $\eta = 0.14$ minor changes of this concurrence dependence are observed (Fig. 1b). The concurrence dependence on $\theta$ has also two maximums and the maximum values, $C = 0.23$, of the concurrence achieves at one point: $\theta = 0.94$ and 2.20 and $\varphi = 0$ and 3.14. A further increase of the asymmetry parameters leads to changes of the dependence on the angles $\theta$ and $\varphi$ (Fig. 1c): several maximums and minimums appear. At $\eta = 0.92$ the maximum concurrence value, $C = 0.35$, is observed at $\theta = 0.40$ and 2.74 and $\varphi = 0$, and $\pi$. The maximum concurrence value and its location dependence on the asymmetry parameters and at $\alpha = \beta = 5$ this value increases from 0.22 up to 0.36 when this parameter grows from 0 till 1.

The concurrence dependence on the magnetic field, quadrupole coupling constant, and temperature is qualitatively independent of the angles. As example, we present below the concurrence as a function of the parameters $\alpha$ and $\beta$ for the concurrence maximum for a spin in the five-coordinated copper ion site ($\eta = 0.14$) at $\theta = 0.94$ and $\varphi = 0$ (Fig. 2). The concurrence increases with the magnetic field strength and inverse temperature and reaches its maximum value. Then the concurrence decreases with increasing the magnetic field strength (Fig. 3). Another dependence of the concurrence on temperature is observed (Fig. 4). At a high temperature concurrence is zero. With a decrease of temperature below a critical value the concurrence monotonically increases till a limiting value. The critical temperature and limiting value are determined by a ratio of the Zeeman and quadrupole coupling energies, $\alpha/\beta$.

## IV. DISCUSSION AND CONCLUSIONS

The obtained results show that the entangled states can be generated between the states of a single nuclear spin $\frac{3}{2}$. From a point of of view of quantum information processing the considered system is isomorphic to a system consisting of two dipolar coupling spins $\frac{1}{2}$. The same quantum logical gates can be realized using the both systems. Therefore the obtained entanglement can be considered as entanglement between qubits formed by states of a single particle. On the other hand, a single spin 3/2 isomorphic to a system consists of two dipolar coupling spins 1/2 and entanglement between the states of a spin 3/2 can be considered as entanglement between two effective spins 1/2. The behavior of entanglement of a spin 3/2 is very similar to that for the system consisting of two dipolar coupling spins $\frac{1}{2}$ [38]. It was obtained that in zero magnetic field the states of the both spin systems, two spins $\frac{1}{2}$ and spin $\frac{3}{2}$, are in separable states. These systems become entangled with increasing the magnetic field. Then, with a further increase of the magnetic field the spin states of the both systems tend to a separable one.

It has recently been shown that, in a system of nuclear spins $s = 1/2$, which is described by the idealized XY model and dipolar coupling spin system under the thermodynamic equilibrium conditions, entanglement appears at very low temperatures $T \approx 0.5 \div 0.3$ $\mu$K [38, 39]. In a non-equilibrium state of the spin system, realized by pulse radiofrequency irradiations, estimation of the critical temperature at which entanglement appears in a system of spins $\frac{1}{2}$ gives $T \leq 0.027$ K [40]. The calculation for $^{63}Cu$ in the five-coordinated copper ion site of $YBa_2Cu_3O_{7-\delta}$ at $\alpha/\beta = 1$, $\eta = 0.14$ and $eQq_{zz} = 62.8$ MHz, gives that the concurrence appears at $\beta = 0.6$ (Fig. 4). This $\beta$ value corresponds to temperature $T \approx 5$ mK. This estimated value of critical temperature is by three orders greater than the critical temperature estimated for the two dipolar coupling spins under the thermodynamic equilibrium [38].

To distinguish an entangled state from a separable one, it is important to determine an entanglement witness (EW) applicable to the given quantum system [41, 42]. The determination of EW is one of the main problems of the experimental study of the entangled states. Internal energy [43], magnetic susceptibility [44], magnetization [45, 46] , and intensity of MQ coherences [40, 47, 48] were used as EW in different systems. Figure 5 illustrates the relation between the concurrence and magnetization, $M$, at $\eta = 0.14$, $\theta = 0.94$, $\varphi = 0$, and $\beta = 10$.In this case the relation between the concurrence and magnetization can be fitted by $C = -\frac{M}{1.9}$ up to $\alpha = 1$ where the linear dependence of the magnetization on the external field is disturbed.

Concurrence depends on the orientation between the external magnetic field and PAF axes. At $\theta = 0$ and $\pi$ the states are separable at any conditions. At $\eta = 0.14$ the concurrence reaches its maximum value at $\theta = 0.94$ and $\varphi = 0$ and $\pi$ (Fig. 1). This effect can open a way to manipulate with the spin states by a rotation of the magnetic field or a sample.

In conclusion, performing investigation has shown that entanglement can be achieved by applying a magnetic field to a single spin 3/2 at low temperature. It was shown that the concurrence is well fitted by a linear dependence on the magnetization in the temperature and magnetic field range up to a deviation of the magnetization from Curie's law and, following, the magnetization can be used as an entanglement witness for such systems. The dependence of the concurrence on the orientation of a sample relative to the external magnetic field opens a way to control the entangled state.

---

Figure Captions

Fig. 1 (Color online) Concurrence as a function of the angles $\varphi$ and $\theta$ at $\alpha = 5$ and $\beta = 5$: a) $\eta = 0$; b) $\eta = 0.14$; c) $\eta = 0.92$.

Fig. 2 (Color online) The maximum concurrence as a function of the parameters $\alpha$ and $\beta$ at $\eta = 0.14$, $\theta = 0.94$, $\varphi = 0$.

Fig. 3 (Color online) Concurrence vs. magnetic field at $T = $ const for various quadrupole interaction constants: black solid line $- \beta = 2$; red dashed line $-\beta = 6$ ; green dotted line $- \beta = 8$; blue dash-doted line $- \beta = 12$.

Fig. 4 Concurrence as a function of temperature at $\frac{\alpha}{\beta} = 0.5$ (black solid line), $\frac{\alpha}{\beta} = 1$ (red dashed line), and $\frac{\alpha}{\beta} = 2$ (blue dotted line) at $\eta = 0.14$, $\theta = 0.94$, $\varphi = 0$ Temperature is given in units of $\frac{eQq_{ZZ}}{4I(2I-1)k_B}$.

Fig. 5 (Color online) Concurrence (black solid line) and magnetization (red dashed line) as functions of the magnetic field at $\beta = 10$, $\theta = 0.94$. Blue dotted line is ( -M/1.9 ).

(a)

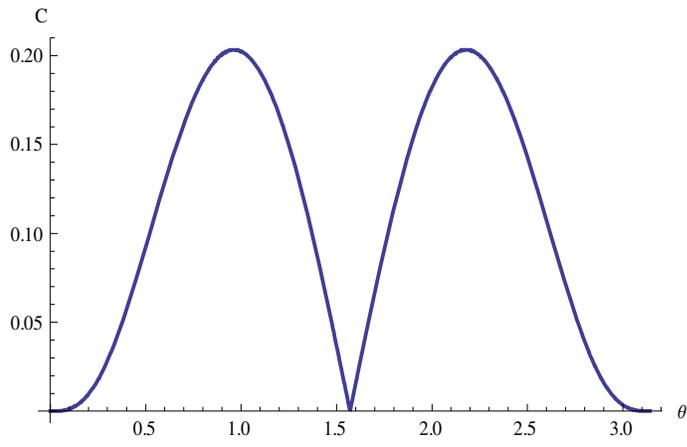

(b)

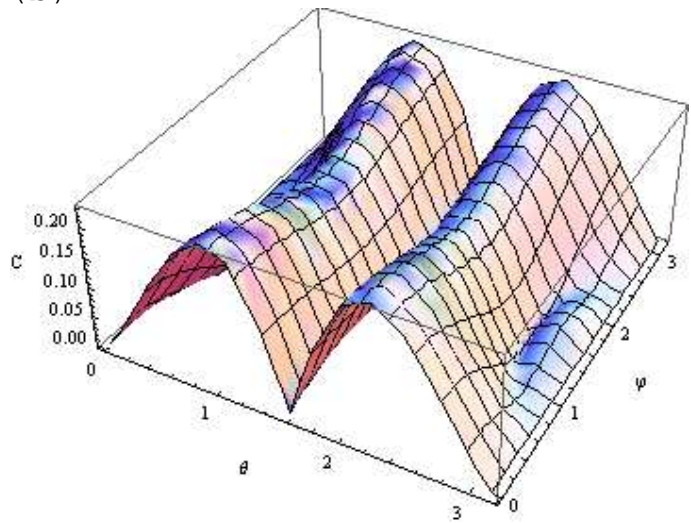

(c)

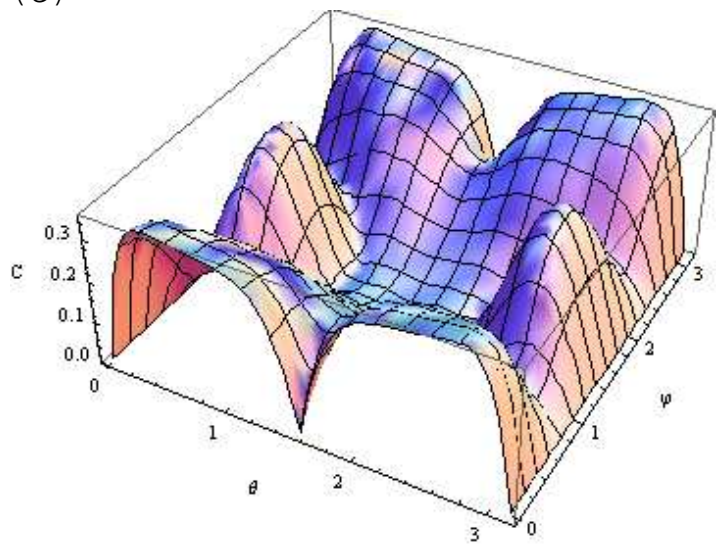

**Figure 2**

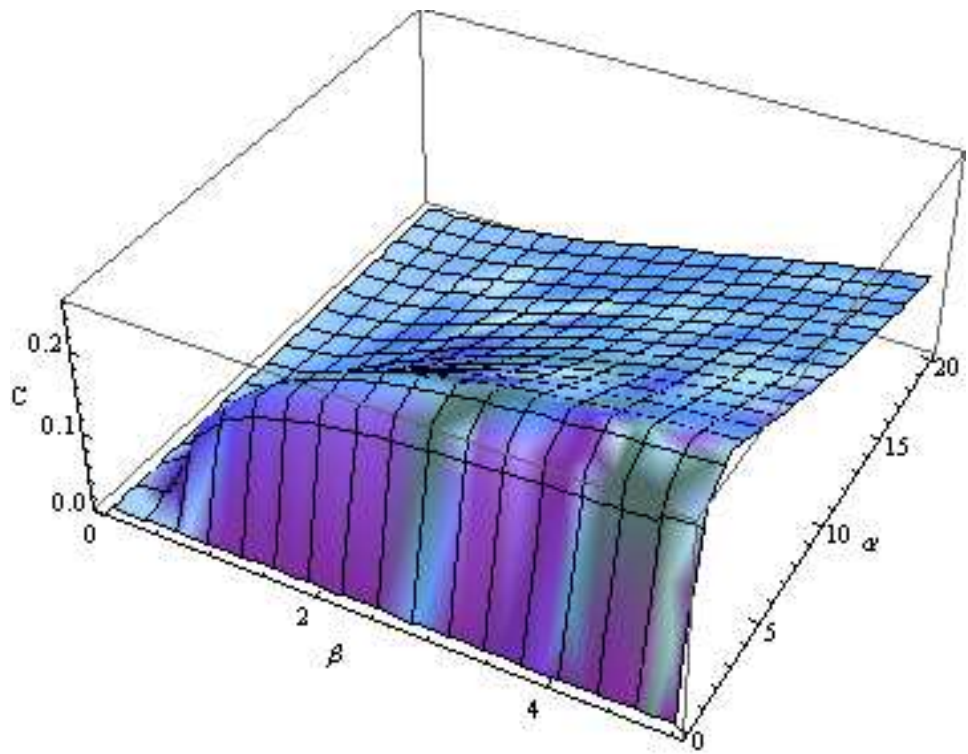

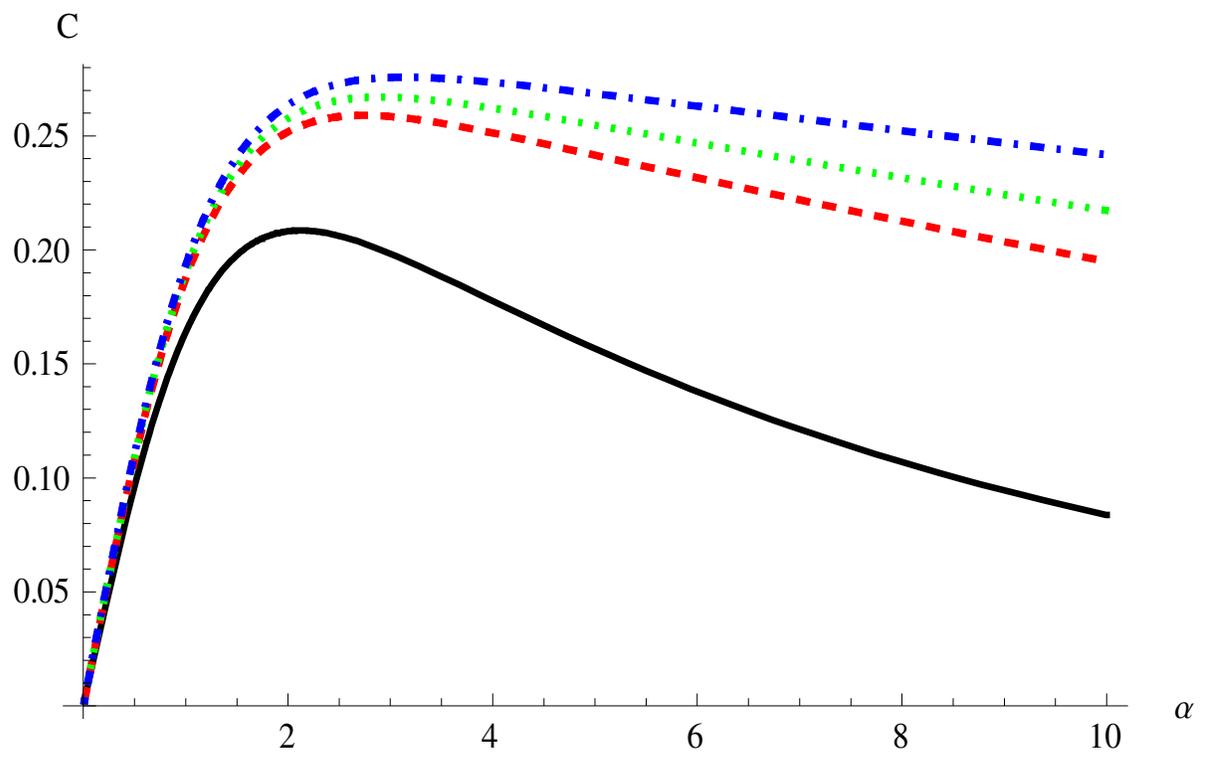

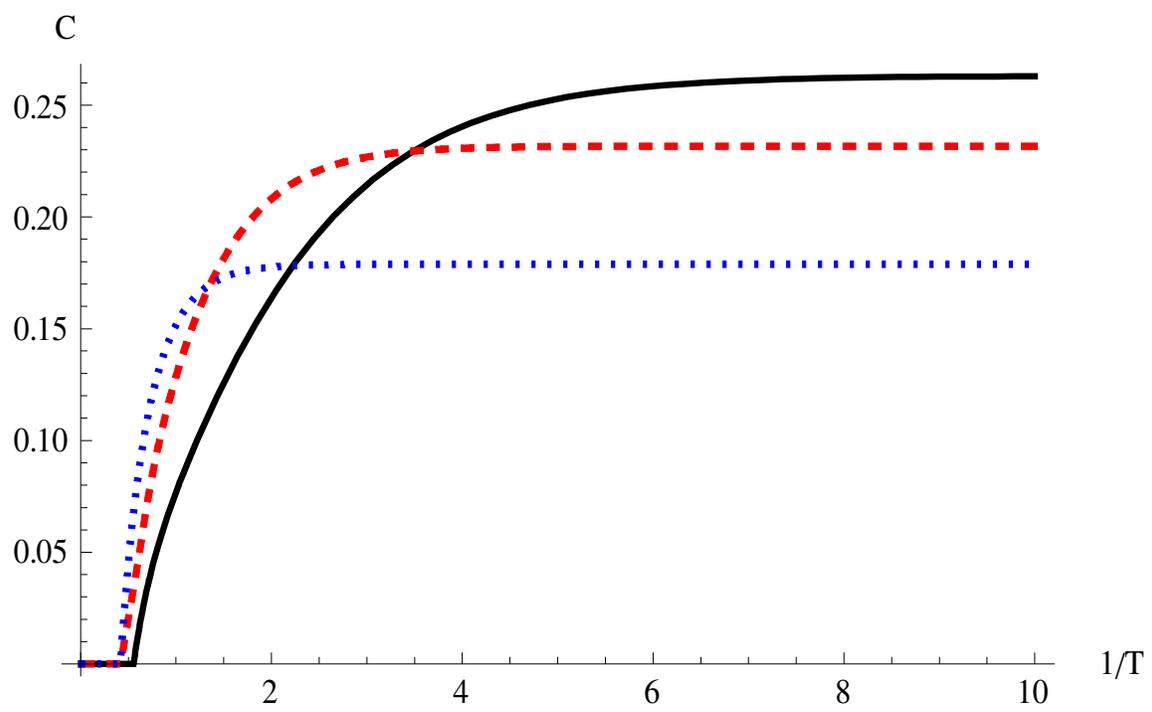

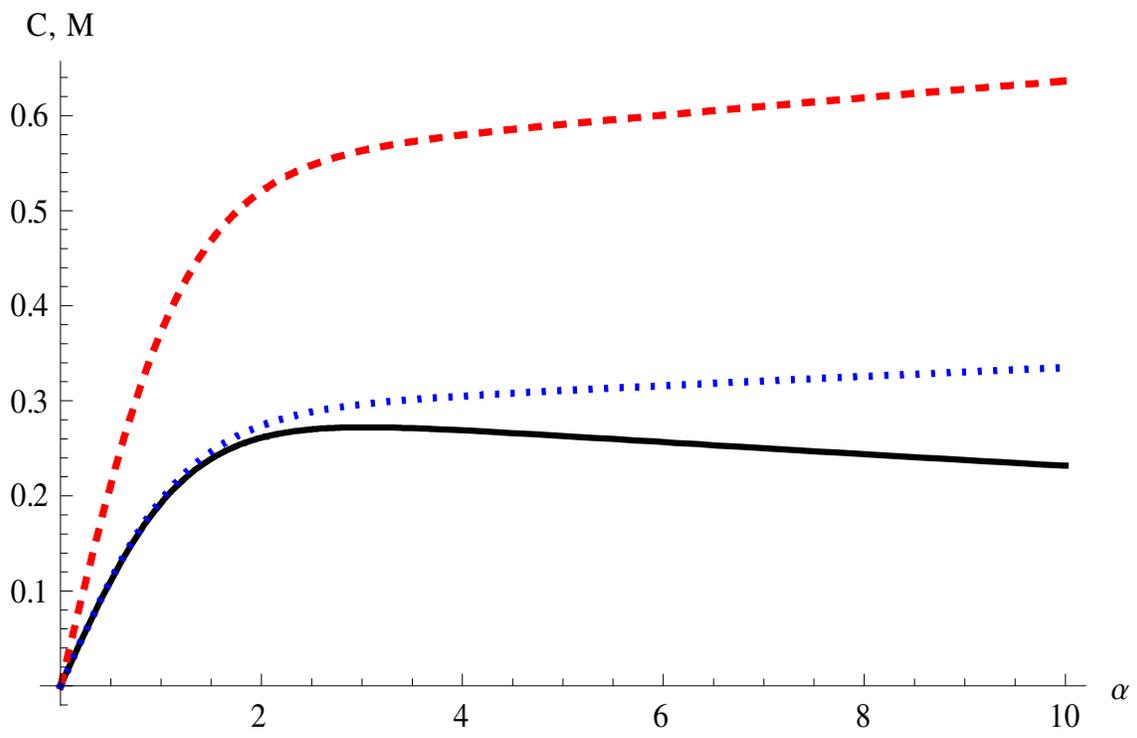